\RequirePackage{fix-cm}
\documentclass[smallextended]{svjour3}          
\usepackage{graphicx}
\usepackage{latexsym}

\usepackage{pstricks}
\usepackage{pst-plot}
\usepackage{pstricks-add}
\usepackage{epsf}

\begin{document}
\title{Sun flux variation due to the effects of orbiting planets. Case of study of a non-compact planetary system}
\titlerunning{Non-compact systems light curve}        
\author{H. Barbier \and E. D. L\'opez \and B. Tip\'an \and C. L. V\'asconez}
\institute{H. Barbier \at  
                 Departamento de F\'isica, Facultad de Ciencias, Escuela Polit\'ecnica Nacional, Quito, Ecuador \\
                 \email{hugo.barbier@epn.edu.ec}
    \and   
                 E. D. L\'opez \at  
                 Departamento de F\'isica, Facultad de Ciencias, Escuela Polit\'ecnica Nacional , Quito, Ecuador\\
                 Observatorio Astr\'onomico de Quito, Escuela Polit\'ecnica Nacional, Quito, Ecuador
   \and      B. Tip\'an \at  
                 Departamento de F\'isica, Facultad de Ciencias, Escuela Polit\'ecnica Nacional , Quito, Ecuador
   \and              
                C. L. V\'asconez \at 
                Departamento de F\'isica, Facultad de Ciencias, Escuela Polit\'ecnica Nacional, Quito, Ecuador 
            }

\date{Received: date / Accepted: date}
\maketitle

\begin{abstract}
We study the phase curves for the planets of our Solar System; which, is considered as a non-compact planetary system. We focus on modeling the small variations of the light curve, based on the three photometric effects: reflection, ellipsoidal, and Doppler beaming.

Theoretical predictions for these photometric variations are proposed, as if a hypothetical external observer would measure them. In contrast to similar studies of multi-planetary systems, the physical and geometrical parameters for each planet of the Solar System are well-known. Therefore, we can evaluate with accuracy the mathematical relations that shape the planetary light curves for an external fictitious observer.

Our results suggest that in all the planets of study the ellipsoidal effect is very weak, while the Doppler beaming effect is in general dominant. In fact, the latter effect seems to be confirmed as the principal cause of variations of the light curves for the planets. This affirmation could not be definitive in Mercury or Venus where the Doppler beaming and the reflection effects have similar amplitudes. 

The obtained phase curves for the Solar System planets show interesting new features that have not been presented before, so the results presented here are relevant in their application to other non-compact systems, since they allow us to have an idea of what it is expected to find in their light curves.


\end{abstract}
\begin{itemize}
\item 
\end{itemize}

\section{Introduction}
\label{intro}

Since the discovery of the first exoplanet orbiting a main sequence star, Pegasi-51b, in 1995 \cite{mayor95}, the number of known exoplanets has been continuously increasing. Mostly supported by the data of $Kepler$ mission \cite{burocki10}, this number has raised up to $3725$ in April 2018. Including $613$ multi-planetary systems \footnote{see http://exoplanetarchive.ipac.caltech.edu}. In the NASA exoplanet archive there are $1559$ confirmed planets in multi-planetary systems with two or more planets, and $763$ in systems with more than two planets. This is the case of: Kepler-11, with $6$ known planets \cite{Gelino2014}, TRAPPIST-1 \cite{gillon17}, and KOI-351, with $7$ planets \cite{Schmitt2014,Cabrera2013}. All of these planetary systems have been detected  by the transit method \cite{borucki84}. Some of the multi-planet systems detected with the radial velocity method \cite{mayor95} are: HD-10180, which has an estimated of $6$ to $9$ planets \cite{lovis11,tuomi12}, and HD-219134 with $6$ planets \cite{Gelino2014}. It is expected that in the coming years the number of detected extrasolar planets will rapidly increase, even in well-known planetary systems, thanks to the forthcoming new generation of high sensitive space instruments and sophisticated data-analysis techniques.

It is commonly accepted that the Doppler beaming effect is negligible compared to the reflection and ellipsoidal effects. With this in mind, this work is focused on studying the phase curve of non-compact systems, for which we compute the ``tiny" photometric variations caused by the reflection, ellipsoidal and the Doppler beaming effects. We use the data of the closest non-compact system, i.e. our own Solar System, for which we know the physical and geometrical parameters needed to evaluate the intensity of each effect. Special attention should be paid on non-transit configurations, which is the usual configuration for a non-compact system. So far, constrained by the methods of detection, most of the discovered multi-planetary systems are very compact, in which the probability for a single --or multiple-- transits increases with the compactness of the system \cite{borucki84}. 

This manuscript is organized as follows. Section \ref{sec1} briefly reviews the astronomical characteristics of the Solar System. In Section \ref{sec2}, we present the relationships that model the reflection, ellipsoidal and Doppler beaming effects, for the eight planets of the Solar System, considering  a fictitious observer located outside the system and in the ecliptic plane. The results and the data analysis are summarized in Section \ref{sec6}. Section \ref{sec7} is dedicated to a final discussion and the conclusions.

\section{The Solar System }
\label{sec1}
In 2006, the International Astronomical Union (IAU) passed resolution B5; which, declared that the Solar System is composed only by eight planets: Mercury, Venus, Earth, Mars, Jupiter, Saturn, Uranus, and Neptune. These are celestial bodies that fulfill the new definitions of planet. We have not considered in these analysis trans-Neptunian objects or small Solar System bodies, due to their negligible effects on the global system dynamics.

Table \ref{tab:1} presents the values of the parameters used to evaluate the phase curves for the planets of the Solar System. Complementary, Table \ref{tab:2} includes some average physical values related to the Sun. Data from Tables \ref{tab:1} and \ref{tab:2} reflect the fact that our planetary system is clearly a non-compact planetary system.

Despite of its main characteristics, the Solar System could easily be compared with some compact systems, e.g. with Kepler-11, which has $6$ planets concentrated in a radius $\sim 0.250$AU around their host star \cite{Gelino2014}, or with TRAPPIST-1, with $7$ planets within a radius of $\sim 0.063$AU from their central star \cite{gillon17}. Kepler multi-planetary system can be considered to have a quasi-circular orbit \cite{Lissauer2012}. In fact, an important eccentricity in a multi-planetary system is generally due to the presence of a hot Jupiter that has enough mass to perturb neighborhood bodies during its migration. These kind of orbital influence is not present in the case of our Solar System. The orbital eccentricities ($e$) of the Solar System planets are relatively weak, except for Mercury ($e \sim 0.206$). Furthermore, the orbits are quasi-coplanar with the only exception, once again, of Mercury. This planet has an orbit inclination of $\sim 7.005^{\circ}$ with respect to the ecliptic plane. On the other hand, in the Solar System, the orbital periods ($P$) vary from $88$ days (Mercury) to $165$ years (Neptune). Meanwhile, in compact multi-planetary systems, $P$ are commonly few days \cite{Gelino2014,gillon17}. Another interesting feature is the location of the habitable zone; it depends on the type of the host star. For dwarf-star planetary systems, the habitable zone is located in the neighborhood of the star. In contrast, in Solar-System types, the habitable zone is further away from its host star \cite{borucki84}.

\section{Theoretical light-curve variation}
\label{sec2}
Recently, ground and space-based telescopes have made it possible to measure and analyze the light curve of different multi-star and multi-planetary systems. The two predominant effects in light-curve profiles have been the --widely known-- reflection and ellipsoidal effects. More recently, the Doppler beaming effect \cite{esteves13} has been measured in observations of eclipsing binaries \cite{hills74}, and exoplanetary systems \cite{Mazeh2012,Faigler2012}. This last effect is recognized to be very weak in intensity with respect to the other ones \cite{loeb03,Zucker2007}, and before the $Kepler$ era it was not detected by space-based instruments. The high sensitivity of the $Kepler$ photometers, let us discriminate the tiny contribution of this photometric effect (down to $10^{-4}$ in magnitude).

\subsection{Planetary Doppler-beaming effect}
\label{sec3}
In 2003, Loeb \& Gaudi \cite{loeb03} first suggested a new description of extrasolar-system light curves; taking into account the Doppler beaming effect (DBE). In their work it is pointed out that it would be possible to measure this effect using the small variability of the flux in $Kepler$-like telescopes \cite{rybicki79}. For this effect, the observed normalized flux variability ($\Delta F / F_0$) oscillates in time-scales near the planetary orbital period $P$ as 
\begin{equation}
  \frac{\Delta F}{F_0} = A_d \sin (\phi),
  \label{eq:5}
\end{equation}
\noindent
where $F_0$ is the mean intrinsic stellar flux, $A_d$ is the amplitude for the DBE, and $\phi$ is the phase angle. We set $\phi = 0$ for the inferior conjunction, and $\phi = \pi$ for the superior conjunction. $A_d$ is related to a bolometric effect due to the Lorentz transformations of the radiated energy. In particular, for the case of non-relativistic velocities, the observed flux $F$ is proportional to the radial velocity $V_r$ \cite{Zucker2007,rybicki79}, 
\begin{equation}
  F = F_0 \left( 1 + 4 \frac{V_r}{ c} \right), 
  \label{eq:1}
\end{equation}
\noindent
where $c$ is the speed of light. Moreover, if the emission is isotropic in the source rest frame, the flux transforms in the same way as the specific intensity, and follows a power-law spectrum. Then, if we consider the band-pass centered on ${\nu_0}$, the normalized flux can be expressed as:
\begin{equation}
 \frac{\Delta F}{F_0} = (3 - \alpha_{d}) \frac{V_r}{c}, 
 \label{eq:3}
\end{equation}
\noindent
where $\alpha_{d}$ is the beaming average spectral index around $\nu_0$  \cite{Zucker2007}. From the data of the Sun (see Table \ref{tab:2}), we found $\alpha_{d} = -1.2211$, corresponding to a frequency ${\nu_0} \sim 5.10^{14}$Hz. 

The radial velocity in turn, can be expressed using its semi-amplitude $K$, by $V_r = K \sin(\phi)$, with:  
\begin{equation}
  K = \left( \frac{2 \pi G}{P} \right)^{1/3} \left[ \frac{M_p \sin(i)}{M_J} \right] 
         \left( \frac{M_\ast}{M_\odot}  \right)^{-2/3} (1-e^2)^{1/2};  
 \label{eq:2}  
\end{equation}
\noindent
here, $G$ is the gravitational constant, $i$ is the orbital inclination respect to the plane of observation, and $M_p$, $M_J$, $M_\ast$, and $M_\odot$ are respectively the masses of the planet, Jupiter, host star, and the Sun. For convenience, it is useful to express the latter equation in ppm units,
\begin{equation}
 K = 28.4~ m s^{-1} \left( \frac{P}{1 year} \right)^{-1/3} \left[ \frac{M_p \sin(i)}{M_J} \right] \left( \frac{M_\ast}{M_\odot} \right)^{-2/3} \frac{1}{(1-e^2)^{1/2}}.  
\end{equation}
\noindent

\subsection{Planetary ellipsoidal effect}
\label{sec4}
The periodical deformation of the visible area of a star, from the point of view of a static-observer, is called ellipsoidal effect \cite{Mazeh2012,Faigler2012,Welsh2010}. In general, it is detected when a massive planet, e.g. a hot Jupiter, is orbiting close to its host star producing a gravitational tidal. This effect is maximum when the line of sigh, and the star-planet axis form a perpendicular angle. Then, the fluxes both radiated, and detected by the observer are maxima.

The relative flux variation due to the ellipsoidal effect, oscillates on time-scales of $P/2$,
\begin{equation}
 \frac{\Delta F}{F_{0}} = A_e \cos (2 \phi), 
 \label{eq:7}
\end{equation}
\noindent
where $A_e$ is the amplitude of this variation. In particular, $A_e$ is affected by the linear ($u$), and gravitational ($y$) darkening limbs \cite{esteves13,loeb03}.
\begin{equation}
 A_e = \alpha_{e} \frac{M_p}{M_\ast} \left( \frac{R_\ast}{a} \right)^3 \sin^2(i),   
 \label{eq:6}
\end{equation}
\noindent
where $\alpha_{e}  = 0.15 (15 + u) (1 + y) / (3-u)$, and $a$ is the semi-major axis of the planet orbit. For the Sun, $u = 0.32$ \cite{cox00}, and $y = 0.45$ \cite{loeb03}.

As the previous case, this amplitude can be rewritten in ppm units as follows, 
\begin{equation}
 A_e = 12.8 ~\alpha_{e} ~sin(i) \left( \frac{R_\ast}{a} \right)^3 \left( \frac{M_\ast}{M_\odot} \right)^{-2} \left( \frac{P_{orb}}{day} \right)^{-2}.   
\end{equation}

\subsection{Planetary reflection effect}
\label{sec5}
For the case of the normalized observed flux variability that is related to the radiation reflection and thermal emission from the planet surface, it oscillates on time-scales of $P$ as: 
\begin{equation}
 \frac{\Delta F}{F_{0}} = A_r \cos (\phi),
\label{eq:9}
\end{equation}
\noindent
where $A_r$ is the reflection amplitude. The amplitude $A_r$ is related to the planetary radius $R_p$, and the geometrical albedo $A_{geo}$, 
\begin{equation}
 A_r = A_{geo} \left( \frac{R_p}{a} \right)^2 sin(i);
 \label{eq:8} 
\end{equation}
\noindent
or in ppm units:
\begin{equation}
 A_r = 570 A_{geo} ~sin(i) \left( \frac{M_\ast}{M_\odot} \right)^{-2/3}  \left( \frac{P_{orb}}{day} \right)^{-4/3}   \left( \frac{R_p}{R_J} \right)^2.
\end{equation}
\noindent
Estimations of the $A_{geo}$ value permit to infer some atmospheric or surface characteristics of the planet (e.g. \cite{alberti2017}).

From here the total fractional variability of the flux ($\Delta F / F_0)_{tot}$ can be computed directly, considering each of these three photometric effects,  
\begin{equation}
 \left( \frac{ \Delta F}{F_0} \right)_{tot} = A_d \sin(\phi) - A_e \cos(2 \phi)  - A_r \cos(\phi),
\end{equation}
\noindent
where $\phi = 0$ corresponds to the primary eclipse (transit), and $\phi = \pi$ to the secondary eclipse (occultation). So, when the planet leaves the transit, all three photometric effects give a positive contribution to the variation of the total flux.

\section{Results}
\label{sec6}
\begin{figure}[!ht]
\includegraphics[width=1\textwidth]{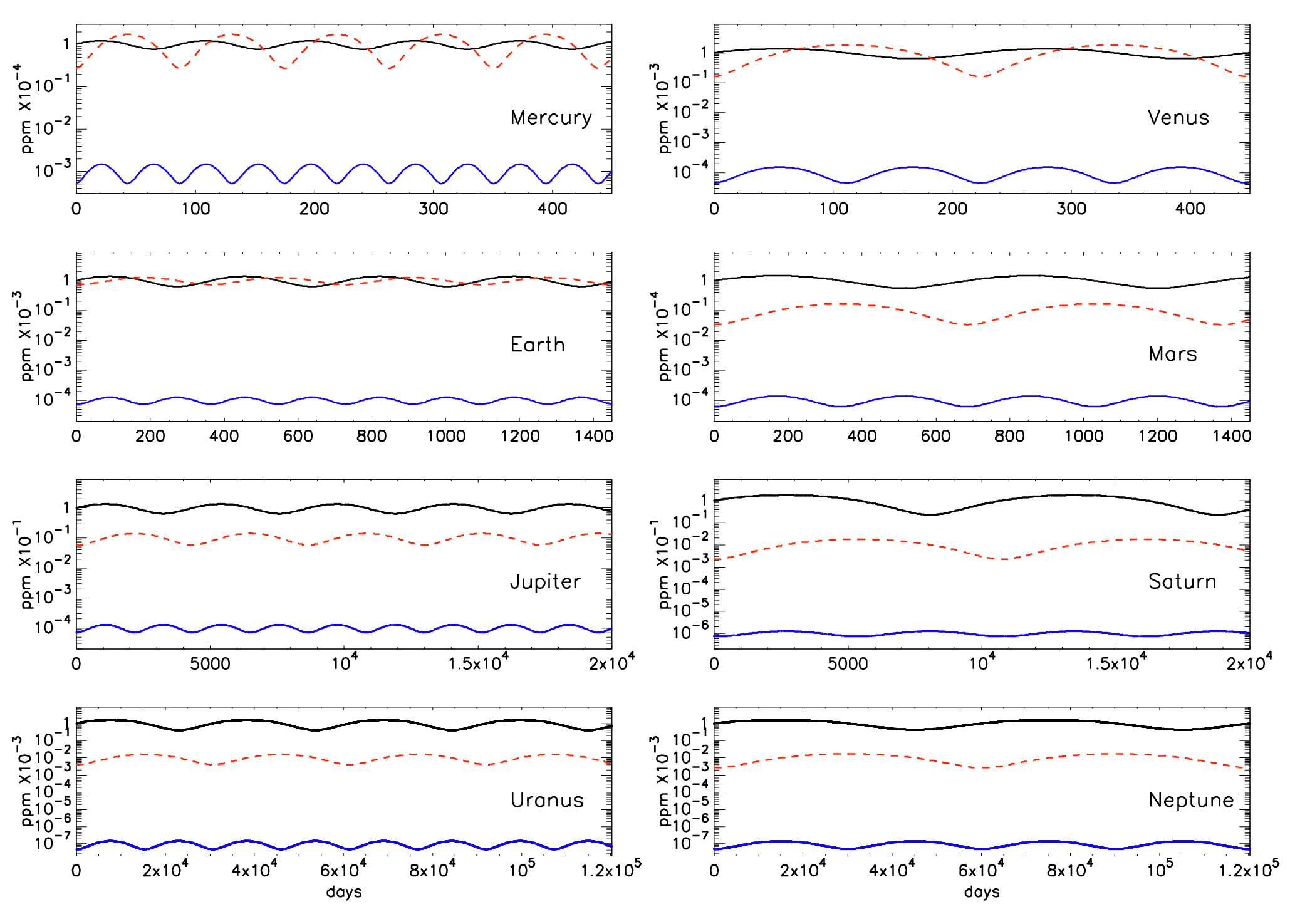}
\caption{Logarithmic relative flux variation ($\Delta F / F_0$) in function of time. Each row has a different time window to improve the visualization. The black solid line corresponds to the DBE, the red dashed line to the reflection effect and the blue solid line to the ellipsoidal effect. The ellipsoidal effect is always the weaker effect in the planets of the Solar System.}
\label{fig:1}       
\end{figure}

In this Section, the theoretical relationships for the three photometric effects of a planetary system are evaluated, for our Solar System. We present in each panel of Fig. \ref{fig:1}, the time-depending contributions of the DBE, reflection and ellipsoidal effects for each planet of the Solar System. We note that the main contribution to the global effect is due to Jupiter, i.e. $1.75\times10^{-7}$ (see Table 3). For all the planets, it is clear to see that the DBE, and the reflection effect are the most important ones. In fact, for the cases of Mercury, Venus, and the Earth these effects are in the same order of magnitude. It is interesting that, in all the planets of the study, the ellipsoidal effect is much weaker than the other effects. Moreover, the longer the distance of the planet to the Sun, the weaker is the contribution of the ellipsoidal effect. This could be considered as one of the criteria of compactness.
We remark that the technology necessary to observe these amplitudes is not yet available. Nevertheless, recent works have successfully studied $Kepler$ space telescope data to model and report planetary phase variations close to the noise level ($\sim 10^{-6}$), e.g. Esteves et al. (2015).

\begin{figure}[!ht]
\includegraphics[width=\textwidth]{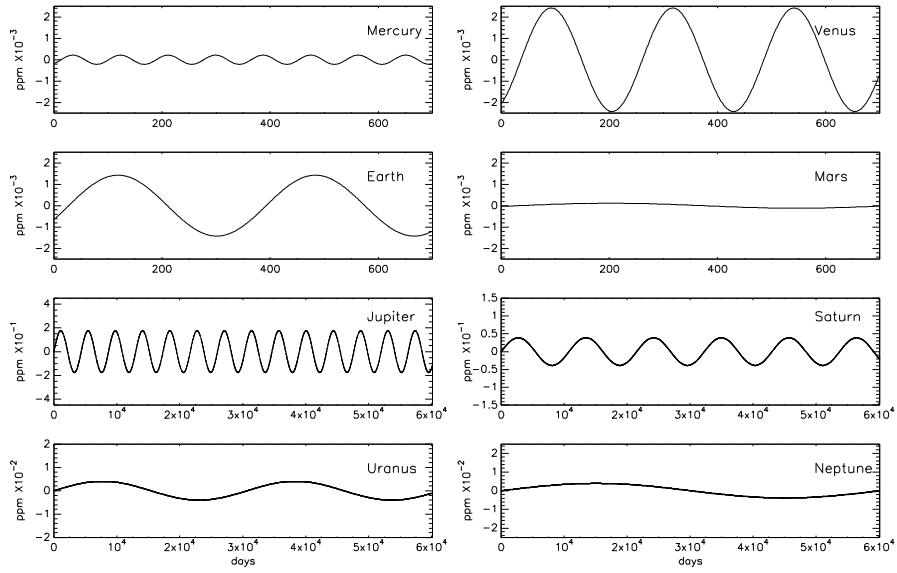}
\caption{The direct sum of the three effects of study that contribute to the total $(\Delta F / F_0)_{tot}$ is computed for each of the Solar-System planets. The time window of the terrestrial planets is one period of Mars. One period of Neptune is used for the rest of the planets.}
\label{fig:2}       
\end{figure}

The panels of Fig. \ref{fig:2} report the accumulative contribution of the three effects to the phase curve, for each of the eight solar planets. The curve in each panel is the direct sum of the three effects, plotted in time windows of one Mars period for the terrestrial planets, and one Neptune period for the rest of the planets. The direct comparison of this composite phase curve let us confirm that the DBE effect is the predominant effect. Except, for the case of the inner planets, where the reflection effect has the same order of magnitude compared with the DBE.

To better highlight these important points, Table \ref{tab:3} shows the maximum values of $A_d$, $A_e$, and $A_r$, for each planet. In general, we note that the DBE is higher than the reflection effect, with the exception of the inner planets: Mercury and Venus. In the latter two cases, the reflection dominates over the beaming effect, but they are almost of the same order. 

\begin{figure}[!ht]
\begin{center}
\includegraphics[width=0.5\textwidth]{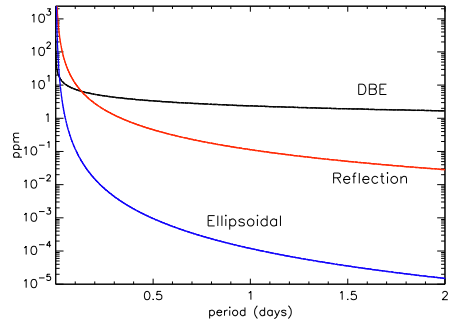}  
\caption{The decay of the amplitude of each effect is plotted with respect to the period of a Jupiter-type planet. The black line represents the DBE ($A_d$), the red line is the reflection effect ($A_r$) and the blue line is the ellipsoidal effect ($A_e$).}
\label{fig:3} 
\end{center}
\end{figure}

For compact systems, it is expected that the reflection, and ellipsoidal effects dominate over the DBE. However, as it can be seen in Fig. \ref{fig:3}, for the example of a Jupiter-type planet, after $\sim 2\times10^{-2}$days, the beaming effect falls down slower ($\sim a^{-1/2}$) than the other effects. At the same time, the reflection effect falls as $\sim a^{-2}$, and the ellipsoidal effect as $\sim a^{-3}$. Then, it is reasonable to suppose that for a non-compact system, like the Solar System, the ellipsoidal effect will be weak for the outer planets. 

The results of this work suggest that the variations of the total flux of the Sun due to orbiting planets is on the order of $10^{-7}$ or less; which, is unobservable by the current technology. However, this constraint could change in the future, or at least the precision will improve a bit from coming missions like PLATO (ESA-SCI (2017)1).

\section{Discussion and conclusion}
\label{sec7}
In this study, we have evaluated the phase curve for the eight planets of the Solar System. The tiny variations of the stellar flux due to the Doppler beaming, reflection, and ellipsoidal effects have been studied from the point of view of a fictitious observer located outside of this non-compact system, but aligned with the ecliptic plane. The individual contribution curves were analyzed to better understanding the entire physical picture of this kind of non-compact systems. 

As the detection of multi-planetary systems is a less-complicated task for large planets located close to their host stars, exoplanets discoveries occur mainly in compact planetary systems; which, have ellipsoidal and reflection photometric amplitudes greater than the Doppler beaming effect.

So far, the studies developed around exoplanets do not take into account the stellar-magnitude fluctuations smaller than  $\sim 10$ppm \cite{esteves13}. Issue directly related to the sensitivity limitations of the currently available space instruments. The $Kepler$ instrument achieved an impressive photometric precision of about 50ppm over six hours for a 12 magnitude target, but it is not enough to detect the tiny amplitudes at the level predicted in our contribution, for the solar system planets. Even future missions like CHEOPS has to measure photometric signals with a precision limited by stellar photon noise of 150ppm per min for a 9th magnitude or 10ppm for stars in 6 hours of integration time \cite{Broeg2013}. In general, for brightest stars, the precision is limited by systematic noise floor of 60ppm per hour. In the case of the recently launched TESS mission, the photometric precision for a 10th magnitude star is estimated to be about 200ppm in 1 hour \cite{Ricker2014} and PLATO (ESA-SCI(2017)1) will reach 27ppm in 1 hour for a 11th magnitude star and 10ppm for a 6th magnitude star.

These photometric signals fall in the background noise and the probability to get results (bumps) from the background requires an accuracy level of 5-sigma to ensure that it is not the result of a statistical fluctuation. On the other hand, we have the restriction on the operation time of space instruments, they have been designed to operate no longer than a few years (4 to 6 years for PLATO), while planet orbits like Jupiter require about 11 years to orbit the Sun and produce at least one whole light curve. In this context, even future instruments are not able to detect extremely low signals as the ones we predicted for the Solar System bodies. Then, the community efforts need to be focused beyond the comprehension of the oscillations of the star and the influence of the stellar spots, active regions, granulation, and activity in general rather than the short-time instrumental evolution \cite{Hippke2015}.

One of the most important conclusions of this study is that for an hypothetical observer outside the Solar System, the Doppler beaming effect is the most relevant, and therefore is the simplest one to be detected, especially for the exterior planets. The contribution of the reflection effect is appreciable, and dominant for the inner planets, while the ellipsoidal effect is practically negligible for the eight planets of our system. These results could be considered as characteristics for a non-compact system.

Despite the non-detectability of such tiny signal amplitudes from the planets of the Solar System, this work is useful to help us qualitatively characterize a non-compact system and facilitates the understanding of its photometric properties, providing some insights for similar studies of other planetary systems.

\bibliographystyle{spphys}       
\bibliography{barbieretal2017}   

\clearpage

\begin{table}[!ht]
 \caption{Data of the Solar System Planets \cite{cox00}.}
 \label{tab:1}       
 \begin{tabular}{lccccccc}
  \hline\noalign{\smallskip}
  Planet  & $P$ ($yr$) & $e$  & $i$ &  $M_p$ ($M_j$)  &  $R_{p}$ (AU)  & $a$ (AU) & $Ag$ \\
  \hline\noalign{\smallskip}
  Mercury  & $2.41\times10^{-1}$ & $2.06\times10^{-1}$	& $83.00$ & $1.73\times10^{-4}$	& $1.63\times10^{-5}$	& $3.87\times10^{-1}$	& $0.10$ \\
  Venus  & $6.15\times10^{-1}$ & $6.77\times10^{-3}$	& $86.61$ & $2.56\times10^{-3}$	& $4.05\times10^{-5}$	& $7.23\times10^{-1}$	& $0.67$ \\
  Earth  & $1.00$ & $1.67\times10^{-2}$ & $90.00$ & $3.14\times10^{-3}$	& $4.26\times10^{-5}$	& $1.00$	        & $0.37$ \\
  Mars  & $1.88$ & $9.34\times10^{-2}$ & $88.15$ & $3.36\times10^{-4}$	& $2.27\times10^{-5}$	& $1.52$	        & $0.15$ \\
  Jupiter  & $1.19\times10^{1}$	& $4.84\times10^{-2}$	& $88.69$ & $1.00$	        & $4.67\times10^{-4}$	& $5.20$	        & $0.52$ \\
  Saturn & $2.94\times10^{1}$	& $5.42\times10^{-2}$	& $87.52$ & $2.99\times10^{-1}$ & $3.89\times10^{-4}$	& $9.58$	        & $0.47$ \\
  Uranus  & $8.40\times10^{1}$	& $4.72\times10^{-2}$	& $89.23$ & $4.40\times10^{-2}$ & $1.70\times10^{-4}$	& $1.92\times10^{1}$	& $0.51$ \\
  Neptune  & $1.65\times10^{2}$	& $8.59\times10^{-3}$	& $88.23$ & $5.35\times10^{-2}$ & $1.65\times10^{-4}$	& $3.01\times10^{1}$	& $0.41$ \\
  \noalign{\smallskip}\hline
 \end{tabular}
\end{table}

\begin{table}[!ht]
 \caption{Physical characteristics of the Sun \cite{cox00}.} 
 \label{tab:2}       
 \begin{tabular}{cccc}
  \hline\noalign{\smallskip}
  $T_\odot$ ($K$) &  $R_\odot$ (AU)  & $u$ & $y$ \\
  \hline\noalign{\smallskip}
  $5778$ & $4.65 \times 10^{-3}$ & $0.32$ & $0.45$ \\
  \noalign{\smallskip}\hline
 \end{tabular}
\end{table}

\begin{table}[!ht]
 \caption{Light-curves maximum variation of the Solar System planets.}
 \label{tab:3}      
 \begin{tabular}{lccc}
  \hline\noalign{\smallskip}
  Planet & $A_d$ & $A_e$ & $A_r$ \\
  \hline\noalign{\smallskip}
  Mercury & $1.13\times10^{-10}$ & $3.51\times10^{-13}$  & $1.81\times10^{-10}$ \\
  Venus   & $1.20\times10^{-9}$  & $8.06\times10^{-13}$  & $2.10\times10^{-09}$ \\
  Earth   & $1.26\times10^{-9}$  & $3.75\times10^{-13}$  & $6.66\times10^{-10}$ \\
  Mars    & $1.09\times10^{-10}$ & $1.13\times10^{-14}$  & $3.32\times10^{-11}$ \\
  Jupiter & $1.75\times10^{-7}$  & $8.46\times10^{-13}$  & $4.19\times10^{-09}$ \\
  Saturn  & $3.87\times10^{-8}$  & $4.05\times10^{-14}$  & $7.76\times10^{-10}$ \\
  Uranus  & $4.02\times10^{-9}$  & $7.39\times10^{-16}$  & $3.96\times10^{-11}$ \\
  Neptune & $3.89\times10^{-9}$  & $2.34\times10^{-16}$  & $1.23\times10^{-11}$ \\
  \noalign{\smallskip}\hline
 \end{tabular}
\end{table}

\end{document}